\documentclass[twocolumn]{aastex631}
\usepackage{amsmath,amstext}
\usepackage{hyperref}
\usepackage{amssymb}
\usepackage{wasysym}
\usepackage{float}
\usepackage{ulem}
\usepackage{breakurl}
\usepackage{color}

	\newcommand{\msun}{\rm M_{\odot}}
        \newcommand{\rsun}{\rm R_{\odot}}
	
	\newcommand{\mbh}{M_{\bullet}}

	\newcommand{\fout}{f_{\text{out}}}

\def\spose#1{\hbox to 0pt{#1\hss}}
\def\lta{\mathrel{\spose{\lower 3pt\hbox{$\mathchar"218$}}
     \raise 2.0pt\hbox{$\mathchar"13C$}}}
\def\gta{\mathrel{\spose{\lower 3pt\hbox{$\mathchar"218$}}
     \raise 2.0pt\hbox{$\mathchar"13E$}}}
 
\righthead{IMBHs in GCs}
\lefthead{Tang et al.}

\journalinfo{Submitted to AAS Journals}

\begin{document}

\title{Searching for Intermediate Mass Black Holes in Globular Clusters Through Tidal Disruption Events}

\author{Vivian L. Tang}
\affiliation{Department of Astronomy \& Astrophysics, University of California, 1156 High Street, Santa Cruz, CA 95064, USA}

\author[0000-0002-6336-3293]{Piero Madau}
\affiliation{Department of Astronomy \& Astrophysics, University of California, 1156 High Street, Santa Cruz, CA 95064, USA}
\affiliation{Dipartimento di Fisica ``G. Occhialini", Università degli Studi di Milano-Bicocca, Piazza della Scienza 3, I-20126 Milano, Italy}

\author[0000-0001-9458-821X]{Elisa Bortolas}
\affiliation{Dipartimento di Fisica ``G. Occhialini", Università degli Studi di Milano-Bicocca, Piazza della Scienza 3, I-20126 Milano, Italy}
\affiliation{INFN, Sezione di Milano-Bicocca, Piazza della Scienza 3, I-20126 Milano, Italy}

\author{Eric W. Peng}
\affiliation{NSF's National Optical-Infrared Astronomy Research Laboratory, 950 North Cherry Avenue, Tucson, AZ 85719, USA}


\begin{abstract}

Intermediate mass black holes (IMBHs) may be the link between stellar mass holes and the supermassive variety in the nuclei of galaxies, and globular clusters (GCs) may be one of the most promising environments for their formation. Here, we carry out a pilot study of the observability of tidal disruption events (TDEs) from $10^3\,\msun<\mbh<10^5\,\msun$ IMBHs embedded in stellar cusps at the center of GCs. We model the long super-Eddington accretion phase and ensuing optical flare, and derive the disruption rate of main-sequence stars as a function of black hole mass and GC properties with the help of a 1D Fokker-Planck approach. The photospheric emission of the adiabatically expanding outflow dominates the observable radiation and peaks in the NUV/optical bands, outshining the brightness of the (old) stellar population of GCs in Virgo for a period of months to years. A search for TDE events in a sample of nearly 4,000 GCs  observed at multiple epochs by the Next Generation Virgo Cluster Survey (NGVS) yields null results. 
Given our model predictions, this sample is too small to set stringent constraints on the present-day occupation fraction of GCs hosting IMBHs. Naturally, better simulations of the properties of the cluster central stellar distribution, TDE light curves and rates, together with larger surveys of GCs are all needed to gain deeper insights into the presence of IMBHs in GCs.

\bigskip
\noindent{\it Unified Astronomy Thesaurus concepts:} Intermediate-mass black holes (816); Globular star clusters (656); Accretion (14).
\end{abstract}

\section{Introduction}

There is currently no unambiguous confirmation for intermediate mass black holes (IMBHs) with masses $10^2\,\msun \lesssim \mbh \lesssim 10^{5}\,\msun$ \citep[for a recent review, see][]{greene2020}. If one were to extrapolate the known $\mbh -\sigma_\star$ relation down to low-mass stellar systems, globular clusters (GCs) would be expected to hosts IMBHs with $\mbh \sim 10^3-10^4\,\msun$ \citep{lutz2013}. Finding objects and characterizing their mass function in this range would provide unique insight into the nature and growth of massive black hole seeds in the early Universe and the dynamical evolution of dense stellar systems, and provide key input
into event predictions for gravitational wave facilities. Thus far there is some circumstantial evidence for black holes below $10^5\,\msun$ in galaxy nuclei (NGC 205, \citealt{nguyen2019}) and in 
hyperluminous X-ray sources (HLX-1, \citealt{webb2012}), but there are still no compelling  candidates with $\mbh\ \sim 10^3\,\msun$.

Different mechanisms have been proposed for the formation of IMBHs in dense stellar clusters, from ``slow" scenarios where the black hole remnant of a massive star sinks to the center of the cluster and grows over time through mergers of mass-segregated lighter black holes \citep{miller2002}, to ``fast" collisional runaway of massive stars during an early phase of cluster core collapse, the product of which may eventually lead to the formation of an IMBH \citep{zwart2002,giersz2015}. It has been shown by \citet{miller2012} that all clusters above a central velocity dispersion of $\sigma_\star \sim 40\,$ km s$^{-1}$ will necessarily form an IMBH through some mechanism at any cosmic epoch, since above this dispersion primordial binaries cannot support the system against deep core collapse. The case is more complicated for lower-dispersion systems such as GCs, where -- regardless of how IMBHs form -- it may be a challenge to retain them in the face of a repeated
onslaught of gravitational wave kicks from mergers with other black holes \citep{holley2008,fragione_kicks2018}.

To date, there are no solid detections of IMBHs in GCs. Claims of dynamical evidence for IMBHs have been made for multiple clusters, but orbital anisotropies and the presence of high concentrations of dark stellar-mass remnants could account for most or all of the alleged IMBH signatures \citep[see][and references therein]{zocchi2019,baumgardt2019,vitral2021}.
IMBH detections have been reported by, e.g., \citet{lutz2013}
for NGC 1904 ($\mbh = 3,000 \pm 1,000\,\msun$) and NGC 6266 
($\mbh = 2,000 \pm 1,000\,\msun$), \citet{feldmeier2013} for 
NGC 5286 ($\mbh = 1,500 \pm 1,000\,\msun$), \citet{ibata2009} for NGC 6715 ($\mbh \simeq  9,400\,\msun$) and 
by \citet{kamann2016} for NGC 6397 ($
\mbh\simeq 600\,\msun$). No radio emission consistent with an accreting $\mbh\gta 1000\,\msun$ IMBH has been observed in a sample of 50 Galactic GCs by \citet{tremou2018}.

To make progress, it is key to develop alternative methods for IMBH searches in dense stellar systems.
Tidal disruption events (TDEs) may offer an independent, promising probe. A TDE occurs when the tidal force of the black hole exceeds the star self-gravity and rips it apart \citep{rees1988}. While about half of the stellar debris is ejected at high speed, the remainder gets accreted,  producing an optical/UV flare accompanied by thermal X-ray emission. One of the most compelling IMBH TDE candidates to date is 3XMM J215022.4--055108
\citep{lin2018}, an X-ray outburst with a luminosity that peaked at $10^{43}$ erg s$^{-1}$ and decayed systematically over 10 years, hosted in a star cluster of mass $\sim 10^7\,\msun$ and plausibly powered by an IMBH of mass 
$\mbh\lta 2.2\times 10^4\,\msun$ \citep{wen2021}. This event suggests that an effective means of detecting IMBHs may be
through a search of optical flares in a large sample of dense star clusters \citep[e.g.,][]{ramirez2009,chen2018,fragione_TDE2018}.
As we shall see, this is particularly true for searches in the NUV, where a disruption event is predicted to outshine the old stellar population of the average GC for a period of months to years.

Large surveys with precise multi-wavelength photometry are becoming increasingly available
for public use. As a pilot experiment, we take advantage of the Next Generation Virgo Cluster Survey (NGVS, \citealt{ferrarese2012}), a comprehensive optical imaging survey of the Virgo galaxy cluster, together with its near-infrared counterpart (NGVS-IR, \citealt{munoz2014}), to set constraints on the fraction of GCs  that may be harboring IMBHs. The NGVS-IR covers a total area of $4\,$deg$^2$ centered on Virgo's core region in the $u^*grizK_s$ bandpasses, and we capitalize on the sample of nearly 4,000 Virgo GCs provided by Peng et al.\ (in prep.). We use theoretical optical light curves to search for potential TDE candidates, and the estimated event rates of TDEs for a given GC population, stellar density profile, and cadence of NGVS observations, to attempt to constrain the fraction of GCs hosting IMBHs with stellar cusps at the present epoch.  While there are considerable uncertainties in the modeling of TDE events around IMBHs in GCs, we hope that our pilot investigation will provide a blueprint for future searches of optical flares in dense stellar systems.

\section{Light curve modeling}
\label{Sec:lc}

While the evolution of a TDE light curve has been extensively modeled in the literature \cite[e.g.,][and references therein]{strubbe2009,lodato2011,strubbe2011,metzger2016,mockler2019,ryu2020,lu2020}, the details of the evolution of the debris stream, the efficiency of the process of circularization, and the emission mechanisms responsible for the optical/UV emission in TDE flares all remain an  open question \citep[for recent reviews see][]{roth2020,rossi2021}. For illustrative purposes, we shall follow here the simplified analysis of \citet{strubbe2009} and \citet{lodato2011}. A star of mass $M_\star=m_\star\,\msun$ and radius $R_\star=r_\star \rsun$ on a very eccentric orbit is torn apart when its pericenter $r_p$ is within the black hole's tidal sphere of radius
\begin{equation}
r_t=\left(\frac{\mbh}{M_\star}\right)^{1/ 3}\,R_\star.
\end{equation}
For lower main-sequence stars, masses and radii are related by 
$r_\star=m_\star^{0.8}$ \citep{kippen1990}. The impact parameter of the encounter, $\beta=r_t/r_p$, measures the strength of the tidal interation.
Initially, approximately half of the initial stellar debris becomes bounded
\citep{evans1989}, and after a fallback time,
\begin{equation}
t_{\rm fb}=\frac{\pi}{\sqrt{2}}
\left(\frac{r_p}{R_\star}\right)^{3/2} \sqrt{\frac{r_p^{3}}{G\mbh}},
\label{eq:tfb}
\end{equation}
most bound material starts coming back to pericenter at the rate
\begin{equation}
\dot M_{\rm fb}={M_\star\over 3 t_{\rm fb}}\left({t\over t_{\rm fb}}\right)^{-5/3}.
\label{eq:Mfb}
\end{equation}
While the early behavior of the fallback rate may be influenced by stellar properties \citep{lodato2009,guillochon2013,law-smith2020},
the late-time accretion rate onto the black hole from a TDE always declines as $t^{-5/3}$ if the star is completely disrupted \citep{coughlin2019}. 

For an IMBH, the mass fallback rate predicted by Equations (\ref{eq:tfb}) and 
(\ref{eq:Mfb}) exceeds the Eddington rate
\begin{equation}
\dot M_E=10 {4\pi G\mbh\over c\, \kappa_{\rm es}}
\label{eq:MEdd}
\end{equation}
for a timescale of years or longer. Here, $\kappa_{\rm es}=0.34$ cm$^{2}$ g$^{-1}$ is 
the electron scattering opacity, and we assume accretion process at $10\%$ radiative efficiency. In this regime, only a fraction of the fallback material joins the newly-formed  Keplerian disk and accretes onto the black hole at the rate
\begin{equation}
\dot M_\bullet=(1-f_{\rm out}) \dot M_{\rm fb},
\end{equation}
while the remaining stellar debris are launched in a radiation pressure-driven outflow during circularization. 
The relative importance of accretion and outflows in this phase is poorly understood \citep[e.g.,][]{ohsuga2005,dotan2011,coughlin2014,metzger2016,wu2018}. To bracket the uncertainties, here we shall adopt two constant values for the fraction $f_{\rm out}$
of infalling gas that is  ejected in a wind, 
$f_{\rm out}=0.1$ \citep[e.g.,][]{strubbe2009,kitaki2021}
and $f_{\rm out}=0.9$ \citep[e.g.,][]{lodato2011,metzger2016}.

\begin{figure}[!h]
\centering
\includegraphics[width=\hsize]{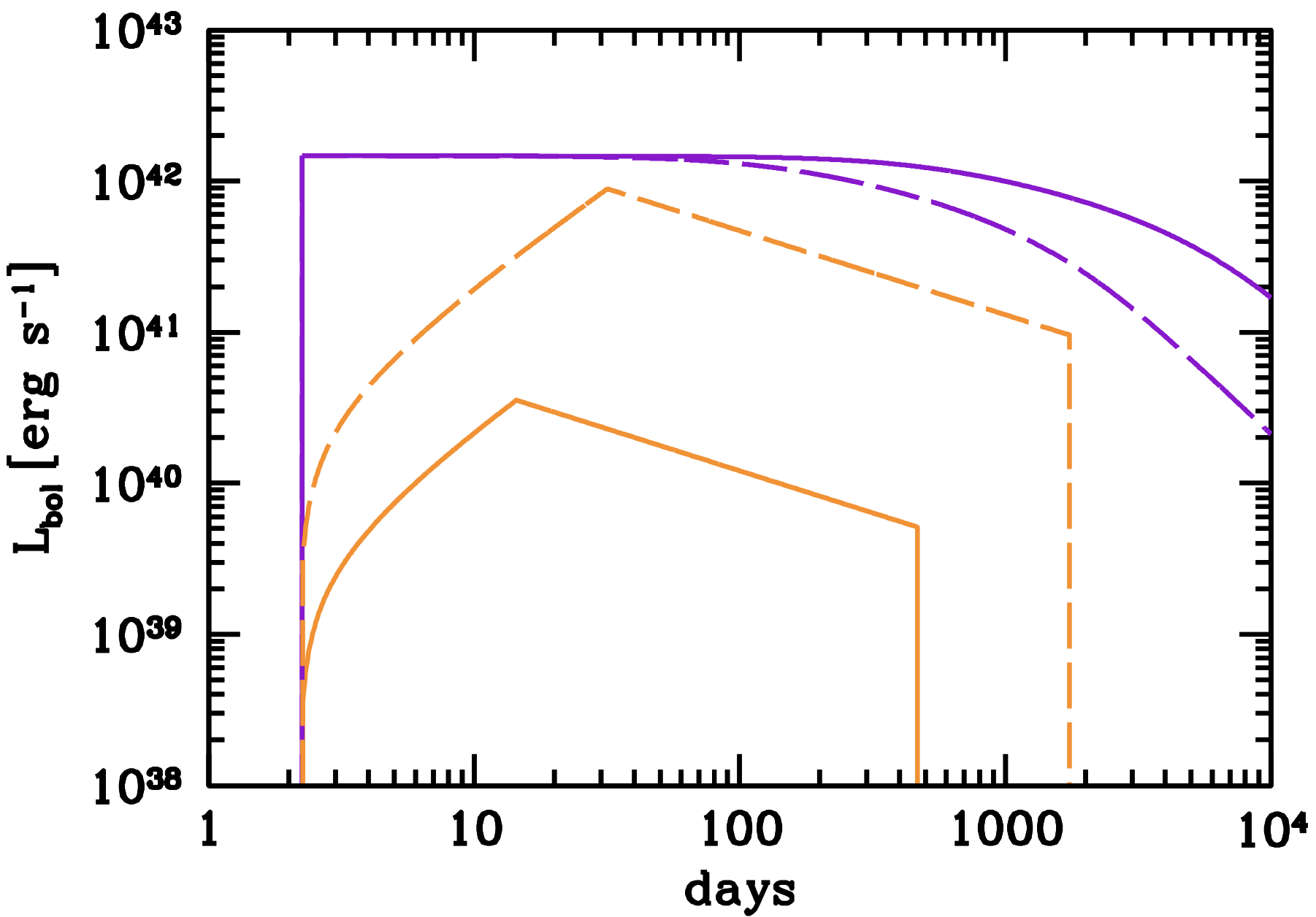}
  \caption{Bolometric light curves for the disk (purple lines) and wind (orange lines) emission resulting from the disruption of a solar-type star by a 3,000 $\msun$ IMBH. The time $t=0$ corresponds to the pericenter passage of the disrupted star, and the fallback time is $t_{\rm fb}=2.24\,$days. The solid and dashed lines show, respectively, the $f_{\rm out}=0.1$ and $f_{\rm out}=0.9$ cases for the fraction of infalling gas ejected in the wind (see text for details). The $f_{\rm out}=0.1$ outflow becomes optically thin at $t=469$ days.
  }
  \label{fig:blum}
\end{figure}

The effective temperature of the 
accretion flow is that of a slim advective disk
\begin{equation}
\begin{split}
\sigma & T_d^4(R)= {3G\mbh\dot M_\bullet f \over 8\pi R^3}\\
& \times \left[\frac{1}{2}+\left\{\frac{1}{4}+\frac{3}{2} f
\left(\frac{10 \dot M_\bullet}{\dot M_E}\right)^{2}\left(\frac{R_S}{R}\right)^{2}\right\}^{1/2}\right]^{-1}
\end{split}
\end{equation}
\citep{strubbe2009}, where $f\equiv 1-\sqrt{3R_S/R}$ and $R_S=
2G\mbh/c^2$ is the Schwarzschild radius. The disk extends from 
$3R_S$ to the circularization radius $R_c=2r_p=2r_t/\beta$, and 
the monochromatic disk luminosity as a function of time is that of a multicolour blackbody,
\begin{equation}
L_\nu=4 \pi^{2} \int_{3R_S}^{R_c} B_{\nu}(T_d) R dR.
\end{equation}
When required, our standard choice of parameters for the stellar
properties is $\beta=1$, $m_\star=1$, and $r_\star=1$. Figure \ref{fig:blum} shows the bolometric 
light curve  for the disk emission from the disruption of solar-type star by a 3,000 $\msun$ IMBH. Note how, during the super-Eddington phase, 
viscosity-generated heat does not have sufficient time to be radiated away, and is instead advected into the hole. The disk effective temperature remains then constant with time even as the fallback rate declines, and  the radiated luminosity saturates at a few times $L_{\rm Edd}$.

A simple model for the radiation-driven wind approximates the geometry as spherical and assumes that stellar debris falls back at close to the escape velocity $v_{\rm esc}= \sqrt{2G\mbh/R_c}$ and shocks at $R_c$, converting  kinetic energy into radiation. Photons are trapped in an outflow promptly launched from the 
circularization radius with $\dot M_{\rm out}(t)=f_{\rm out}\dot M_{\rm fb}$ and terminal velocity $v_w=f_v\,v_{\rm esc}$. Our standard choice for the wind velocity parameter is $f_v=1$. The radiation temperature at the base of the wind, $T_c$, can
be derived from energy conservation in the wind
\begin{equation}
4\pi R_c^2 v_w (aT_c^4)={1\over 2}\,\dot M_{\rm out}\,v_w^2.
\label{eq:Tc}
\end{equation}
The outer radius of the ejecta grows as $R_{\rm ej}=R_c+v_w(t-t_{\rm fb})$, and the outflow is optically thick to electron scattering out to a photospheric radius $R_{\rm ph}$ given by $R_{\rm ph}\, \rho_{\rm ph}\,{\kappa}_{\rm es}=1$. 
The gas density profile for $r<R_{\rm ej}$ follows from matter conservation, $\rho(r,t)=
\dot M_{\rm out}(t)/(4\pi r^2 v_w$). 
The temperature of the advected radiation decreases adiabatically as $T
\propto \rho^{1/3}$, and photons escape with a  blackbody spectrum at the photospheric temperature of
\begin{equation}
T_{\rm ph}=T_c\left({\rho_{\rm ph}\over \rho_c}\right)^{1/3},
\end{equation}
where $\rho_c=\dot M_{\rm fb}/(4\pi R_c^2 v_{\rm esc})$ is the gas density at the base of the flow. 

\begin{figure}[!h]
\centering
\includegraphics[width=\hsize]{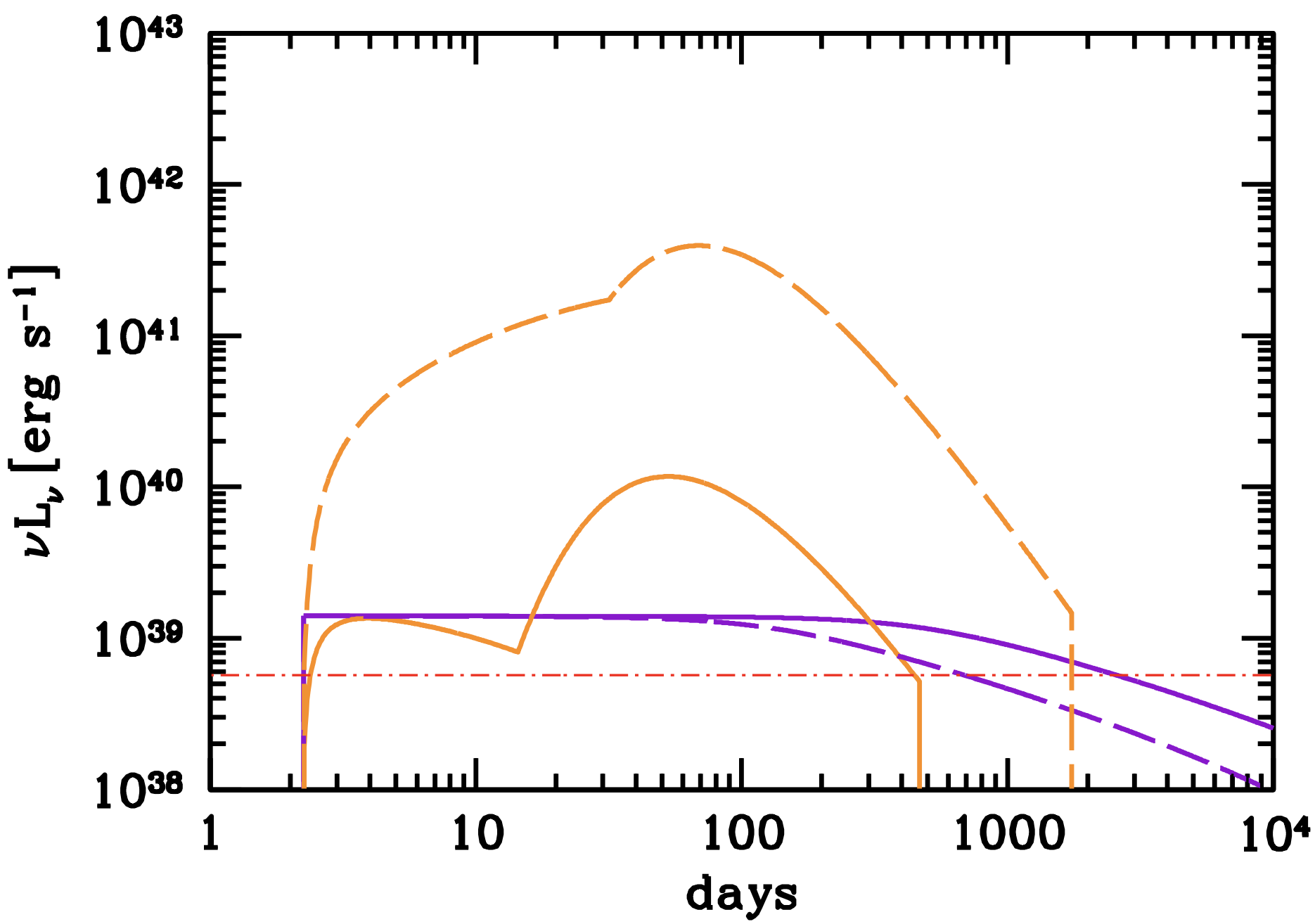}
  \caption{Same as Figure \ref{fig:blum} but in the $u$-band ($\nu=8\times 10^{14}\,$Hz). The red dash-dotted line shows the mean $u$-band luminosity, 
  $\nu L_\nu=5.7\times 10^{38}$
  erg s$^{-1}$ of the 
  GCs detected in the NGVS survey \citep{jordan2007}.
  }
  \label{fig:Ulum}
\end{figure}

At the earliest times, the fallback rate can be so large and the density so high that the location of the photosphere is just inside $R_{\rm ej}$. In this ``edge-limited'' phase, we set $R_{\rm ph}=R_{\rm ej}$ and $\rho_{\rm ph}=(R_{\rm ej}\,{\kappa}_{\rm es})^{-1}$. The specific outflow luminosity 
is given by
\begin{equation}
L_\nu=4 \pi^{2} R_{\rm ph}^2 B_\nu(T_{\rm ph}).
\end{equation}
Figure \ref{fig:blum} shows the resulting wind bolometric light curve, which rises as $t^{11/9}$ during the edge-limited initial phase, when the photosphere expands and $T_{\rm ph}$ decreases as $t^{-7/36}$. As time passes and the density of the outflow subsides, the photosphere sinks inward
as $t^{-5/3}$, its temperature rises as $t^{25/36}$, and the luminosity  declines as $t^{-5/9}$.
The total radiation luminosity of the wind is of order $L_{\rm Edd}$ in the $f_{\rm out}=0.9$ case, decreasing for lower outflow rates as $f_{\rm out}$ during the edge-limited phase, and as 
$f_{\rm out}^{5/3}$ afterwards \citep[cf.][]{strubbe2009}. These relations only apply for $R_{\rm ph}\ge R_c$, because otherwise the outflow is optically thin. We impose this limit by setting the wind luminosity to zero when the photosphere shrinks below $R_c$. In the $f_{\rm out}=0.1$ case the wind becomes optically thin when the fallback rate is still super-Eddington.

While the disk always dominates the total radiated power, the bulk of the disk emission occurs in the soft X-rays. The optical/NUV flash plotted in  Figure \ref{fig:Ulum} is instead produced by the adiabatically expanding outflow, which largely outshines the disk during the optically thick phase. For comparison, we also show in the figure the mean NUV luminosity of GCs detected in the NGVS survey, which argues for the detectability of a TDE optical transient against the brightness 
of a dense star cluster. 

Depending on the uncertain geometry of the outflow and the viewing angle of the observer to the source, it is possible that both the accretion disk and the outflow may be visible at early times. Here, we shall neglect obscuration effects and sum up the two contributions to produce a total TDE light curve.

\section{TDE Rates: Estimates}
\label{Sec:Estimates}

To provide estimates and trends of the number of TDE flares expected in a local GC survey like the NGVS, we present in this section a simple and intuitive configuration-space approach following \citet{syer1999}. Our treatment is informed by the more rigorous numerical integration of the one-dimensional, orbit-averaged 
Fokker-Planck equation for the evolution of the stellar distribution function, which we perform and discuss in the next section.

Stars within a cluster of current mass $M_{\rm GC}$ are assumed to follow a Plummer density profile
\begin{equation}
\rho_\star(r)={\rho_c\over  (1+r^2/a^2)^{5/2}},
\end{equation}
with 1D velocity dispersion
\begin{equation}
\sigma_\star^2(r)={\sigma_c^2
\over \sqrt{1+r^2/a^2}}.
\end{equation}
Here, $\rho_c=3M_{\rm GC}/(4\pi a^3)$ and 
$\sigma^2_c={GM_{\rm GC}/(6a)}$ are the density and velocity dispersion in the cluster core, and $a$ is
the Plummer scale parameter that sets the size of the cluster core. 
We have fit the masses and structural parameter data of Milky Way's GCs of \citet{Baumgardt2018} with the core density-cluster mass relation:
\begin{equation}\label{eq:rho_core}
\rho_c=1.7\times 10^3\,{\msun\over {\rm pc^3}}\,
\left({M_{\rm GC}\over 10^5\,\msun}\right)^{2}.
\end{equation}
For a GC with mass $M_{\rm GC}=5\times 10^5\,\msun$, the expressions above yield $\rho_c=4.3\times 10^4\,\msun\,{\rm pc^{-3}}$, $a=1.4\,$pc, and $\sigma_c=16\,$km s$^{-1}$.

At the center of this Plummer cluster we place an IMBH with $\mbh\ll M_{\rm GC}$. 
Following mass segregation \citep{bahcall1977}, low mass main-sequence stars and their remnants relax to a Bahcall-Wolf $r^{-\gamma}$ density cusp near the black hole, with $3/2<\gamma<7/4$ \citep{alexander2009}. We thus write the stellar density near the center of the cluster as

\begin{equation}
\rho_\star(r)={\rho_c}\,[1+(R_i/r)^{\gamma}],
\label{eq:rhocusp}
\end{equation}
where $R_i\ll a$ is the radius of influence of the black hole. This is defined  as the location where the interior stellar mass equals $\mbh\,$ \citep[e.g.,][]{binney1987,merritt2004,wang2004}, i.e.
\begin{equation}
R_i \equiv 
\left[g(\gamma){\mbh \over M_{\rm GC}}\right]^{1/3}a,
\end{equation}
where $g(\gamma)=(3-\gamma)/ (6-\gamma)$.
In our set-up, $R_i$ is typically a few times larger than the radius $r_\bullet\equiv {G\mbh/\sigma_c^2}$ defined kinematically in terms of  the intrinsic stellar velocity dispersion. For IMBHs weighing a few percent of the  cluster core mass, collisional $N$-body simulations show that the $\rho_\star\propto r^{-\gamma}$ cusp  extends all the way up to $R_i>r_\bullet$ \citep{baumgardt2004}. 
This is also confirmed in our direct numerical integration of the Fokker-Plank equation for the time-dependent stellar distribution in the vicinity of an IMBH (see next Section). 
The Jeans equations associate an $r^{-\gamma}$ density cusp with a velocity dispersion that near the black hole approaches $\sigma_\star^2\rightarrow G\mbh/[(1+\gamma)r]$
\citep[e.g.,][]{zhao1996}. We therefore approximate the cluster central velocity dispersion as
\begin{equation}
\sigma_\star^2(r)=\sigma_c^2+{GM_\bullet\over (1+\gamma) r}.
\end{equation}
Hereafter, we shall assume a slope of $\gamma=7/4$ for the central stellar cusp.

\begin{figure}[!b]
\centering
\includegraphics[width=\hsize]{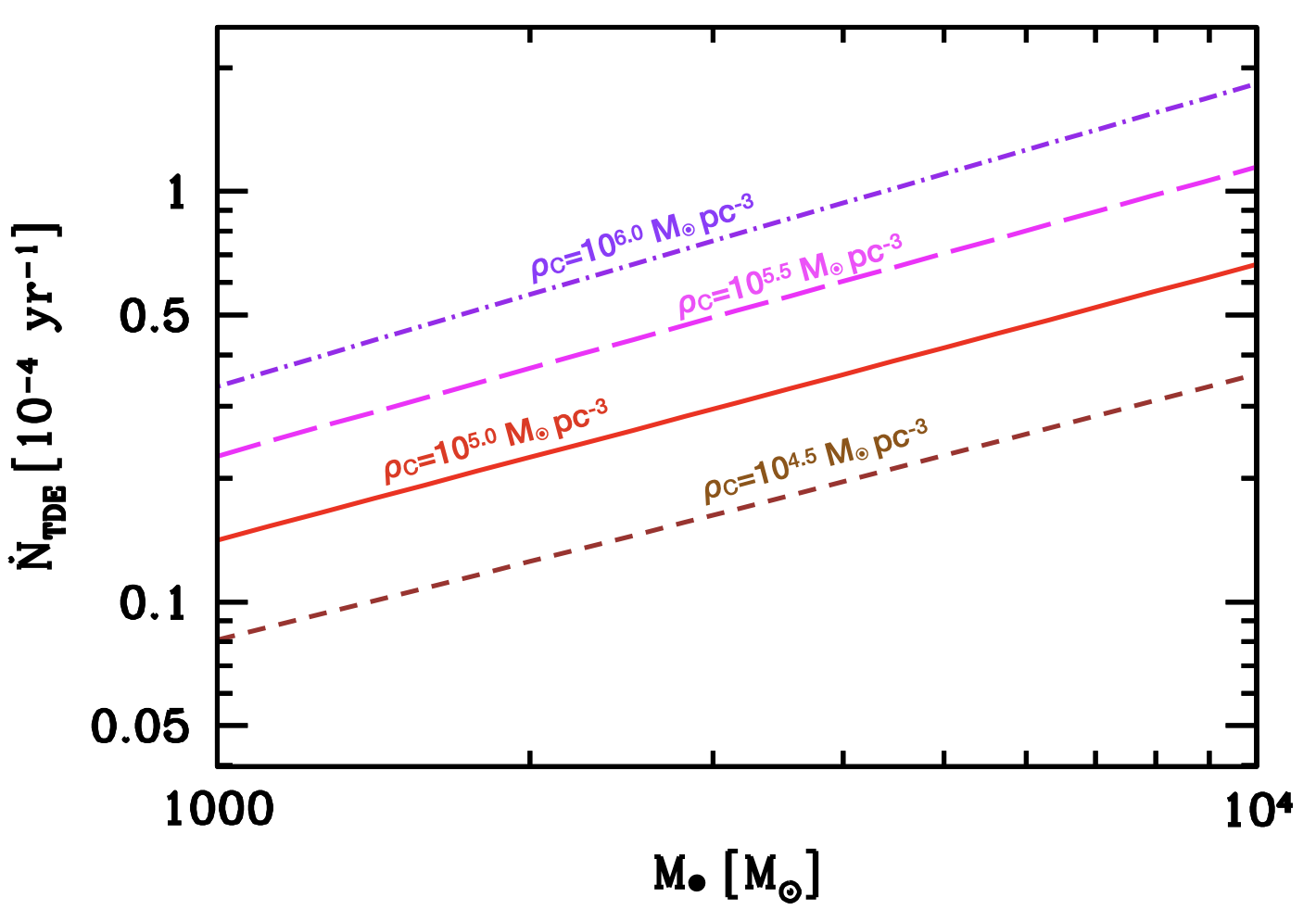}
  \caption{Predicted TDE rates for $10^3\,\msun<\mbh/\msun<10^4$ IMBHs embedded in $r^{-7/4}$ stellar cusps at the center of GCs with 
  $\rho_c=10^{4.5}\,\msun\,$pc$^{-3}$ (short-dashed curve), 
  $10^{5.0}\,\msun\,$pc$^{-3}$ (solid curve), 
  $10^{5.5}\,\msun\,$pc$^{-3}$ (long-dashed curve), and
  $10^{6.0}\,\msun\,$pc$^{-3}$ (dot-dashed curve).
  Both the empty and full loss-cone regimes contribute an $\mathcal{O}(1)$ fraction of the total TDE rate. These calculations assume a single-mass stellar system composed of $M_\star=1\,\msun$ stars. 
  }
  \label{fig:TDErates}
\end{figure}
\begin{figure*}[t]
\centering
\includegraphics[width=0.8\hsize]{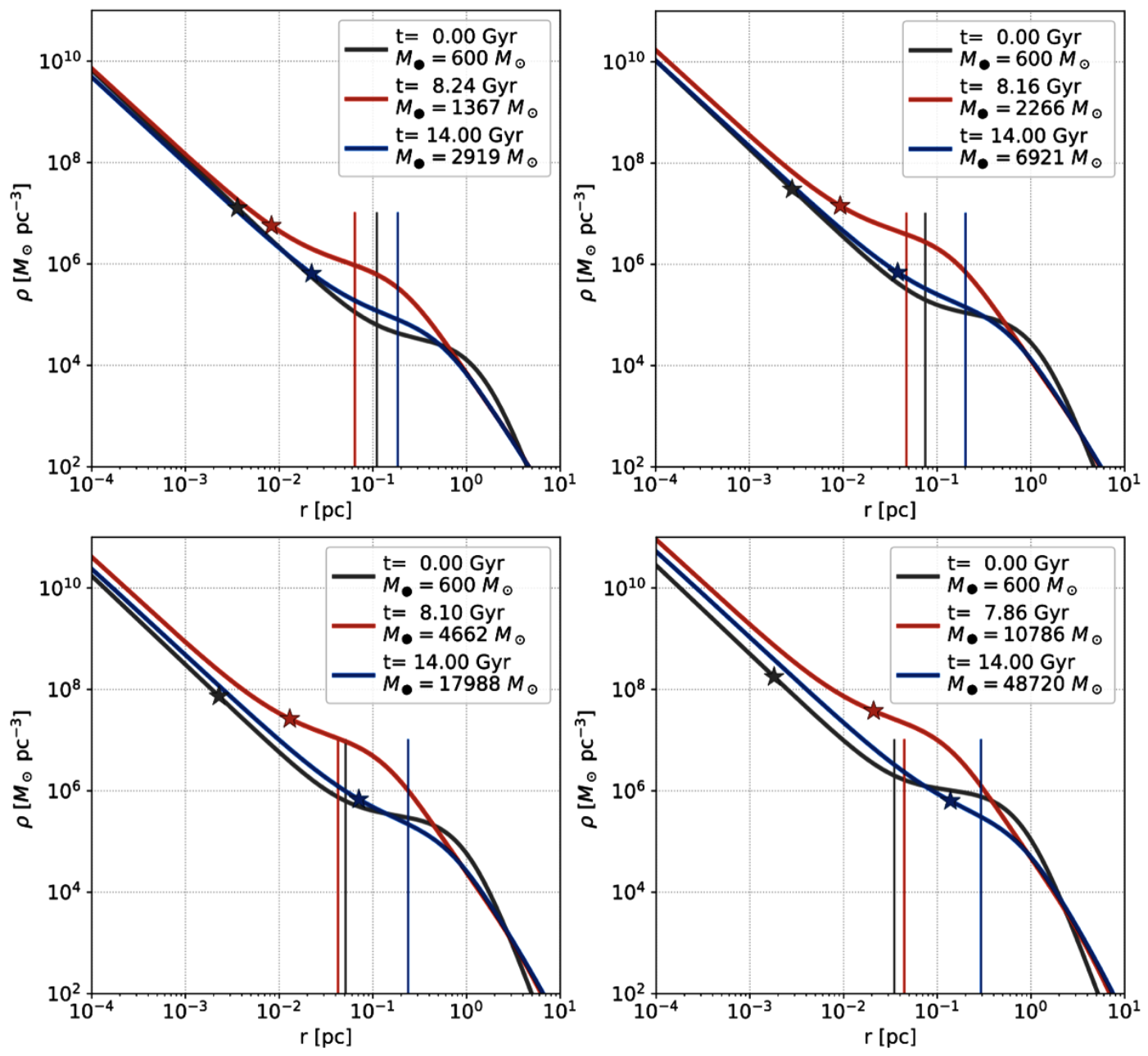}
  \caption{Evolution of the stellar density profile of four IMBH$+$GC systems with initial Plummer core densities
  $\rho_c=10^{4.5}\,\msun\,$pc$^{-3}$ (top-left panel), 
  $10^{5.0}\,\msun\,$pc$^{-3}$ (top-right panel), 
  $10^{5.5}\,\msun\,$pc$^{-3}$ (bottom-left panel), and
  $10^{6.0}\,\msun\,$pc$^{-3}$ (bottom-right panel). In 
  all cases, a central IMBH of initial mass $\mbh(t=0)= 600\,\msun$ is embedded in a pre-existing $r^{-7/4}$ stellar cusp that extends all the way up to the influence radius (vertical lines in the panels). The starred points denote the locus where most TDEs are sourced. The collisional evolution of these spherical isotropic stellar systems under two-body relaxation and loss-cone effects was simulated by integrating the coupled Poisson and orbit-averaged 1D Fokker-Planck equations with the code {\sc PhaseFlow} \citep{vasiliev2017} for a single-mass stellar system composed by $M_\star=1\,\msun$ stars.
  }
  \label{fig:FP_profiles}
\end{figure*}

Stars on nearly parabolic orbits are tidally disrupted when their specific  angular momentum $J$ is smaller than $J_{\rm lc} = (2G\mbh r_t)^{1/2}$. The ensemble of nearly-radial orbits with $J<J_{\rm lc}$ forms the so-called ``loss cone", the set of velocity vectors at some distance $r$ from the black hole that lie within a cone of half-angle $\theta_{\rm lc}$ \citep{frank1976,lightman1977}
\begin{equation}
\theta_{\rm lc}^2  = {G\mbh r_t\over \sigma_\star^2r^2}.
\end{equation}
Stars in the loss cone are disrupted at the first periapsis passage, and the continued supply of stars to the IMBH 
is driven by gravitational deflections that repopulate
the loss cone on a timescale  that is related to the two-body relaxation time
\begin{equation}
t_R={0.34\,\sigma^3_\star\over G^2 \langle M_\star^2\rangle n_\star \ln\Lambda}
\label{eq:tR}
\end{equation}
\citep{spitzer1987}. Here,  $n_\star =\rho_\star/\langle M_\star\rangle$ is the 
number density of main sequence stars and their remnants,  and we take $\ln \Lambda=\ln (0.4\,\mbh/\langle M_\star\rangle)$ for the Coulomb logarithm. In a simplified treatment, two dynamical regimes can be defined depending on the ratio of the dynamical timescale, $t_d=r/\sigma_\star$, to the diffusion timescale of angular momentum across the loss cone, $t_{\rm lc}=\theta_{\rm lc}^2t_R$. Close to the IMBH, orbital periods are short and stars diffusing into the loss cone are immediately disrupted. This is the empty loss-cone regime,  and the rate of TDEs per star can be estimated as 
\begin{equation}
{d\dot N_<\over dN_{\rm MS}}={1\over \ln(2/\theta_{\rm lc})t_R},
\label{eq:dotNE}
\end{equation}
where $N_{\rm MS}(<r)$ is the number of 
main-sequence stars contained within a radius $r$ \citep{frank1976,lightman1977,syer1999}.
This flux is proportional to the relaxation rate but only weakly dependent on the size of the loss cone. In the other regime, $t_d>t_{\rm lc}$, scattering in and out of the loss cone is faster than the orbital time, and the loss cone will 
always be full, uniformly populated in orbital phase.
The fraction of stars in the loss cone at any time is then just $\theta_{\rm lc}^2$, and the TDE rate per star is given by \citep{syer1999}
\begin{equation}
{d\dot N_>\over dN_{\rm MS}}={\theta_{\rm lc}^2\over t_d},
\label{eq:dotNF}
\end{equation}
independent of the stellar encounter timescale. After solving for the transition radius $r_{\rm crit}$ where the two per-star disruption rates in Equations (\ref{eq:dotNE}) and 
(\ref{eq:dotNF}) are equal, we write  the total TDE rate of main-sequence stars as 
\begin{equation}
\begin{split}
\dot N_{\rm TDE} =\dot N_< + &  \dot N_>  =  ~4\pi \int_0^{r_{\rm crit}}
{n_{\rm MS}\,r^2\over \ln(2/\theta_{\rm lc})t_R} dr \cr
 + & ~4\pi \int_{r_{\rm crit}}^\infty
{n_{\rm MS}\, G\mbh r_t\over \sigma_\star r}dr.
\end{split}
\end{equation}
%

%
In the above equations, 
the term $\langle M_\star\rangle$ represents the change in the event rate with the number of stars at fixed total stellar density $\rho_\star$, while the second moment $\langle M_\star^2 \rangle$ is responsible for the decrease of the diffusion timescale as the gravitational potential becomes more ``granular" and stellar-mass black holes dominate the relaxation rates
\citep{kochanek2016,bortolas2022}. In general, to compute $\langle M_\star\rangle$ and
$\langle M_\star^2\rangle$ one needs to adopt a present-day mass function (PDMF) 
with an upper truncation at $M_\star^{\rm max}\approx 1\,\msun$ to approximate the   old stellar population of a GC \citep{magorrian1999}. For simplicity, and in analogy with the time-dependent Fokker-Planck numerical approach discussed in the next section, we will 
assume instead an idealized, single-mass stellar system composed by $M_\star=1\,\msun$ stars ($n_{\rm MS}=n_\star$). Compared to a single-mass stellar population, a realistic PDMF is known to enhance TDE rates by several-fold \citep{stone2016}
even after accounting for the smaller tidal radii, $r_t\propto M_\star^{-1/3}\,R_\star\propto M_\star^{7/15}$, of sub-solar stars.\footnote{In our model, the empty loss-cone disruption rate can be shown to scale as 
$\dot N_<\propto r_t^{4/9}$
up to a slowly varying logarithmic factor.}

%
%

{Figure \ref{fig:TDErates} shows the predicted TDE rates for $10^3<\mbh/\msun<10^4$ IMBHs in GCs with 
a single-mass stellar population and increasing core densities in the range $\rho_c=10^{4.5}-10^6\,\msun\,$pc$^{-3}$.
Most events are sourced at $r_{\rm crit}< R_i$, and both the empty and full loss-cone regimes contribute an $\mathcal{O}(1)$ fraction of the total disruption rate. The rates of TDEs increase with black hole mass and cluster core density approximately as $\dot N_{\rm TDE} \sim (\rho_c\mbh)^{0.5}$, and typically exceed $10^{-5}\,{\rm yr^{-1}}$ for $\gta 10^3\,\msun$ IMBHs harbored in dense GCs.
Obviously, these high rates cannot be sustained indefinitely because of stellar depletion. In the next section we will compute TDE rates in a time-evolving stellar density profile -- including depletion --  using a Fokker-Plank approach.

\section{TDE Rates: Fokker-Planck Integration}

We have tested the simplified model of the previous section against the results of the  {\sc PhaseFlow} Fokker-Plank integrator of \citet{vasiliev2017}. The publicly available {\sc PhaseFlow} code solves the coupled Poisson and orbit-averaged 1D Fokker-Planck equations, evolving a spherically symmetric distribution of stars under two-body relaxation and loss-cone effects in the neighbourhood of a central black hole, the mass of which is allowed to grow with time following stellar captures \citep{bortolas2022}. We explore a one-parameter family of models, varying the initial core density of the initial Plummer profile between $10^{4.5}-10^6\,\msun\,{\rm pc}^{-3}$ and adjusting the total mass of the cluster according to Equation (\ref{eq:rho_core}) ($M_{\rm GC} = 4.3\times10^5 - 2.4\times10^6\,\msun$). In all configurations, a central IMBH of initial mass $\mbh(t=0)=600\,\msun$ is embedded in a $r^{-7/4}$ initial stellar cups that extends all the way up to the influence radius $R_i$. Following each capture event, 30\% of the mass of the disrupted star is accreted by the IMBH. For simplicity, we assume here an idealized, single-mass stellar system composed by $M_\star=1\,\msun$ stars, postponing a numerical treatment that includes a complete, time-dependent stellar mass function and the effect of mass segregation on TDE rates to future work.

Figure \ref{fig:FP_profiles}  shows the evolving stellar density profiles of four IMBH$+$GC systems having different initial core densities in the range $\rho_c= 10^{4.5}-10^6\,\msun\,$pc$^{-3}$ (as in Figure \ref{fig:TDErates}). As shown by \citet{vasiliev2017}, 
the cusp is not a steady-state structure: the gray line shows the initial Plummer+Bahcall–Wolf cusp profile ($t=0$), the red line corresponds to the time when the cusp amplitude attains its maximum value, and the blue line depicts the profile at the end of the integration, $t=14\,$ Gyr, when the cusp density normalization has decreased in response to the heat source at the center. In the figure, the vertical lines mark the influence radius $R_i$ containing a mass of stars equal to $\mbh(t)$, while the starred points denote the locus where most TDEs are sourced. 

The  evolution  of all clusters follows  a  similar  route. At first the system expands, powered by an outward heat conductive flux driven by two-body relaxation. The density in the cusp decreases, roughly maintaining an $r^{-7/4}$ profile inside the influence radius. The initial expansion phase lasts $\sim 0.1\,$Gyr, which is of order the two-body relaxation timescale at the influence radius
$R_i(t=0)$, and is followed by a phase of secular gravo-thermal contraction leading to a maximum cusp density at time $t_{\rm peak}\simeq 8\,$ Gyr; as expected, the latter timescale is of order the relaxation timescale measured in the outer regions of the cluster. IMBH-star interactions eventually generate enough heat to prevent further core collapse and cause a late re-expansion of the cluster. 

\begin{figure}[!ht]
\centering
\includegraphics[width=\hsize]{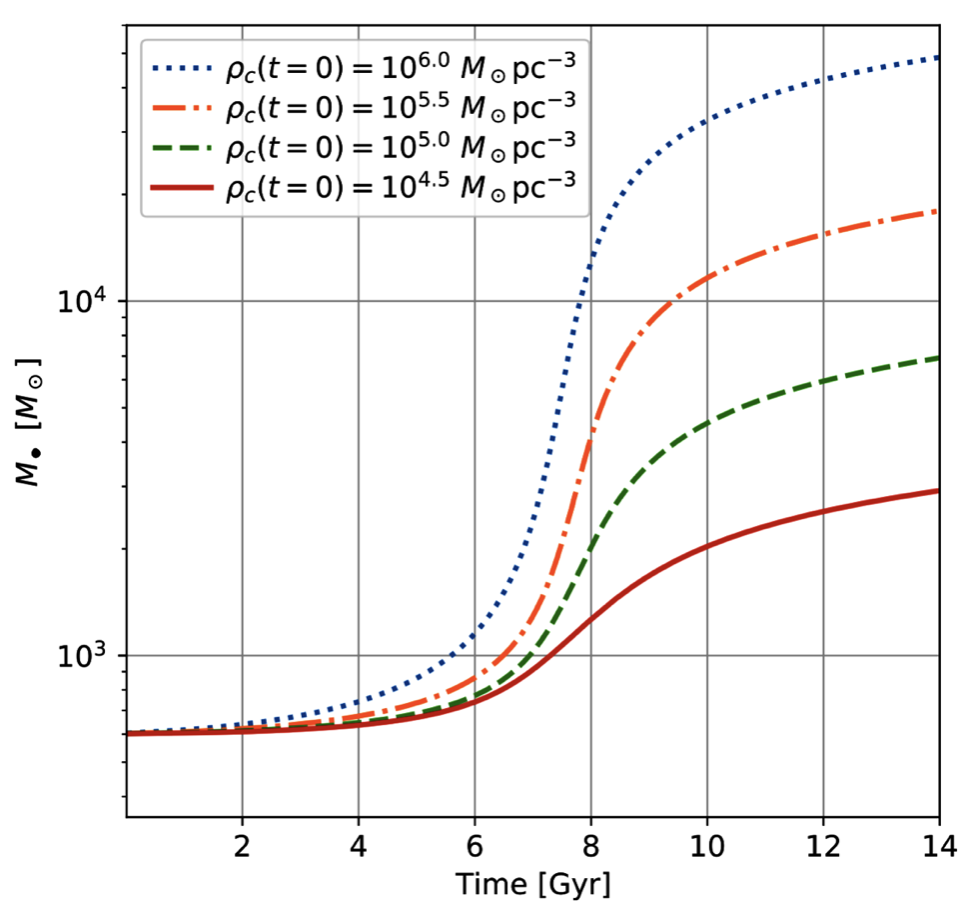}
  \caption{The growth of central IMBHs by stellar captures in the time-evolving stellar density profiles of Figure \ref{fig:FP_profiles}. In all configurations, the initial black hole mass was  fixed to $\mbh(t=0)= 600\,\msun$. Plummer core initial densities range from $\rho_c(t=0)=10^{4.5}\,\msun\,$pc$^{-3}$ (solid line) at the bottom to $10^{6.0}\,\msun\,$pc$^{-3}$ (dotted line) at the top. 
  }
  \label{fig:IMBH_growth}
\end{figure}

During the 14 Gyr evolution, the IMBH acquires a mass of $2,919-48,720\,\msun$ depending on the chosen initial stellar density, with a more efficient growth in the case of denser and more massive systems (Figure \ref{fig:IMBH_growth}) and a faster rise at time 
$t=7-9\,$ Gyr during the gravo-thermal cluster contraction. The three main phases of the GC density profile progression -- fast expansion, contraction of the core, and slow re-expansion -- are reflected in the evolution of the TDE rate (Figure \ref{fig:TDE_FP}), which is simultaneously modulated by the growth of the central black hole. Disruption rates reach a minimum after $0.1-0.2\,$Gyr, then climb dramatically by more than 2 orders of magnitude during core contraction, peak at $t_{\rm peak}\simeq 8$ Gyr, and slowly decay at late times. The peak in the rates is clearly sourced by the collapse of the outer regions of the cluster delivering a substantial amount of new stars near the IMBH and promoting its growth.  Peak rates are more pronounced in initially denser clusters where stellar captures lead to more massive IMBHs.
The orbital radius where most disruption events originate (starred point in  Figure \ref{fig:FP_profiles}) moves outward during the core expansion phases, and inward when the core contracts.

\begin{figure}
\centering
\includegraphics[width=\hsize]{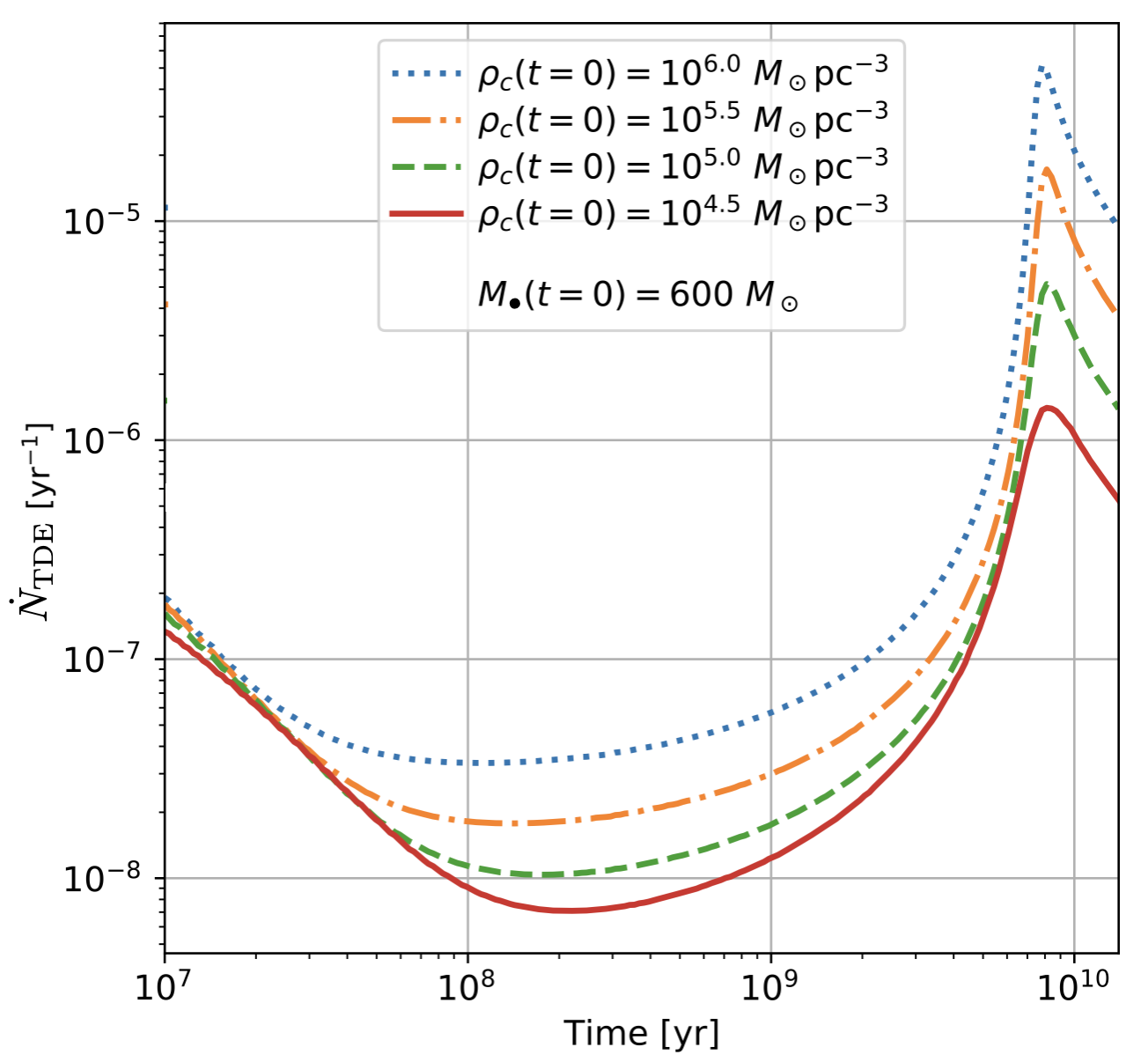}
  \caption{Predicted TDE rates in the time-evolving stellar density profiles of Figure \ref{fig:FP_profiles}. A single-mass stellar system composed by $M_\star=1\,\msun$ stars was assumed for simplicity in the numerical integration of the coupled Poisson and orbit-averaged 1D Fokker-Planck equations with {\sc PhaseFlow}. In all cases, a central IMBH of initial mass $\mbh(t=0)=600\,\msun$ was embedded in a pre-existing $r^{-7/4}$ stellar cusp  (see text for details).
  }
  \label{fig:TDE_FP}
\end{figure}

Because of the intrinsic time-dependence of the problem, a detailed comparison between our steady-state analytical estimates in Section \ref{Sec:Estimates} and the numerical results obtained with the {\sc PhaseFlow} code is far from trivial. The Fokker-Planck integration validates qualitatively:

\begin{itemize}

\item the assumed steady-state inner density profile, which actually evolves in amplitude in Figure
\ref{fig:FP_profiles} but approximately retains the same $r^{-7/4}$ functional form, with little dependence on initial conditions;

\item the IMBH mass range used to estimate light curves and TDE rates in GCs, which is consistent with the mass accreted in $\sim 7-8\,$ Gyr by sufficiently massive black hole seeds (with seed masses above a few hundred solar masses) via stellar captures in dense stellar systems (see Figure\ref{fig:IMBH_growth}); and

\item the trends of increasing TDE rates with $\mbh$ and $\rho_c$ (Figure \ref{fig:TDErates}), which are also seen in our numerical integration model (Figure \ref{fig:TDE_FP}).

\end{itemize}
When compared with our steady-state analytical estimates for a single-mass stellar population of $1\,\msun$ stars (Figure \ref{fig:TDErates}), the peak rates observed in the Fokker-Planck integration appear to be a few times lower than the corresponding rates of the simplified model. 

In our analysis below of the NGVS GC sample, we shall adopt for simplicity the analytical rates of Figure \ref{fig:TDErates}, as they are expected to be intermediate between the numerical results and steady-state estimates that include a complete, evolved stellar mass function. A useful power-law fit to these TDE rates as a function of $\mbh$ and $\rho_c$ is given by
\begin{equation}
\dot N_{\rm TDE} =a\left(\frac{\rho_c}{\mathrm{\msun\,pc^{-3}}}\right)^{b} \left(\frac{\mbh}{10^5\,\msun} \right)^{c},
\label{eq:plfit}
\end{equation}
with parameters $a=(2.2\pm 0.3)\times 10^{-6}\,{\rm yr}^{-1}$,
$b = 0.44 \pm 0.002$, and $c= 0.71 \pm 0.005$.\footnote{Note that such an event rate cannot be sustained indefinitely, as nearly the entire cluster would be consumed  over a Hubble time. The time evolution in Figure ~\ref{fig:TDE_FP}  shows that high rates are only achieved at late times. Considering that most GCs are about 10 Gyr old, i.e. they are close in age to the peak of the TDE rate in the Figure, it is reasonable to use this value as a good estimate for the present-day TDE rate.}

\begin{figure*}[!ht]
\centering  
\vspace{0.cm}
\includegraphics[width=0.8\hsize]{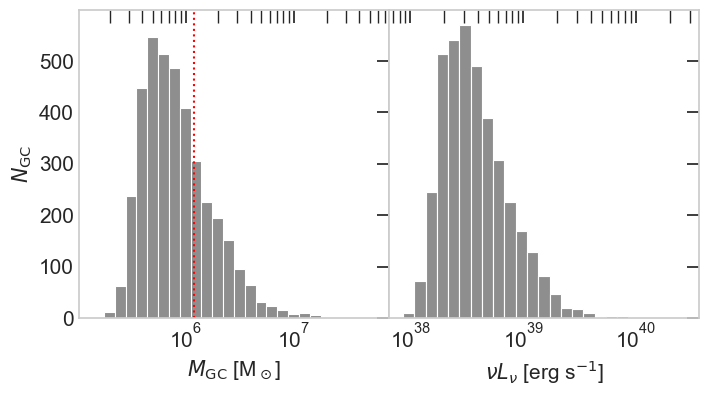}
  \caption{Stellar mass (left panel) and mean monochromatic luminosity (right panel) distributions of NGVS GCs. Stellar masses are computed using the $z$-band mass-to-light ratios predicted by the metallicity-dependent population-synthesis model PEGASE \citep{jordan2007}. The vertical red dashed line marks the mean value of the distribution, $\langle M_{\rm GC}\rangle= 1.1\times10^6\,\msun$. In the $u$-band, most GCs are fainter than $10^{39}$ erg s$^{-1}$.
   }
  \label{fig:ngvs_dist}
\end{figure*}
%


\section{NGVS GC Sample}\label{data}

To compare our theoretical predictions with observations, we search for optical flares from TDEs in a robust sample of GCs in $u$- and $g$-band. These clusters are in the core of the nearby Virgo galaxy cluster within a $2\,$deg$^2$ region centered on M87. The photometric data are part of the NGVS \citep{ferrarese2012}
and its near-infrared counterpart, NGVS-IR \citep{munoz2014}.  GC candidates were selected using extreme deconvolution \citep{bovy2011} to model the distribution of foreground stars, GCs, and background galaxies in a multidimensional parameter space of color and morphology (concentration parameter), which determined the probability of a given source to be a GC (Peng et al., in prep.). Our sample of NGVS globulars consists of CFHT MegaCam time-series photometry in the $ugiz$ bandpasses spanning 5 years with a cadence timescale ranging from hours to months and sometimes several years. The cadence was determined by the need to make large dithers for studying the low surface brightness outskirts of massive galaxies \citep{ferrarese2012}. Most GCs have about 5–20 measurements in each of the four filters with $0.05-0.2$ mag photometric precision depending on the apparent magnitude of the system. After removing those with only single-epoch observations, our final NGVS sample of GCs with $u$-band photometry includes a total of 3,849 sources. 

\subsection{Photometric Variability}

A temporary brightening of a NGVS GC could indicate a potential TDE, a false positive due to cosmic ray hits, or other image artifacts. We have checked for large photometric variations, applying a sigma-clipping technique to find outliers in subsets of data. Four outliers at $3.5\sigma$ were found in the $u-$band, and image inspection showed that cosmic rays were causing the increase in brightness. Our search for detectable TDE events in the NGVS GC sample yields null results.

\subsection{Mass and Luminosities}

Stellar masses of NGVS GCs were computed from their AB magnitudes $m_{\rm AB}$ as
\begin{equation}
m_{\rm AB}=C-2.5 \log M_{\text{GC}}.
\label{eq:Mgc}
\end{equation}
The constant $C$ includes the logarithm of a mass-to-light ratio $\Upsilon$ in any arbitrary band-pass and the distance $d$ to the Virgo cluster [16.5 $\pm{0.1}$ (random) $\pm{1.1}$ (systematic) Mpc, \citealt{Mei_2007}], and is given by
\begin{equation}
C=\msun+2.5 \log \Upsilon+5\log \left(\frac{d}{10}\right).
\label{eq:Mgc_const}
\end{equation}
From the reported MegaCam photometry, we take the average $g$- and $z$-band magnitudes and compute for each GC the color $(\overline{g}-\overline{z})$. The colors are then used to obtain $z$-band mass-to-light ratios by adopting the corresponding values predicted by the metallicity-dependent population-synthesis model PEGASE \citep{jordan2007}. The resulting NGVS GC mass distribution is shown in  Figure \ref{fig:ngvs_dist}. The GC sample has mean stellar mass  $\langle M_{\rm GC}\rangle= 1.1\times10^6\,\msun$, and 31\% of all globulars are more massive 
than $10^6\,\msun$. 
We used $z$-band rather than $g-$band for mass determinations because  mass-to-light ratios at redder wavelengths are less sensitive to the exact age or star formation history of a stellar population, minimizing systematic uncertainties.
Figure \ref{fig:ngvs_dist} also shows the $u$- and $g$-band monochromatic luminosities of our GC sample. Globulars are typically fainter in $u$, with NUV 
luminosities rarely exceeding $10^{39}$
erg s$^{-1}$.

\subsection{Detectability of TDEs in NGVS GCs}\label{method}

To assess quantitatively the observability of TDEs in our GC sample, and in preparation for future multi-epoch surveys of extra-galactic star clusters, we assume here that each NGVS GC hosts an IMBH of mass $0.01\,M_{\rm GC}$.
The input parameters for each cluster are its stellar luminosity, the corresponding TDE rate (Eq. \ref{eq:plfit}), and the theoretical light curves in the two cases, $f_{\rm out}=0.1$ and $f_{\rm out}=0.9$, chosen to bracket the strength of the radiation-driven wind component. For each GC we then compute the time interval $\Delta t_{\rm TDE}$ during which the TDE {\it outshines} the steady stellar GC component in the $u$-band, as well as the total time span between the first and last epoch of cluster observation, $\Delta t_S$. The number of observable TDEs in the sample is then
\begin{equation}
N_{\rm TDE}=f_{\rm occ}\sum_i \dot N_{\rm TDE,i}\,(\Delta t_{\rm TDE,i}+\Delta t_{S,i}). 
\label{eq:NTDE}
\end{equation}
Here, the sum is over all GCs, $f_{\rm occ}<1$ is the occupation fraction, and the first term on the RHS corrects for the events that went off before the 
start of the survey but whose light-curve tail would still be visible at survey times. Note that typically $\Delta t_{\rm TDE}>\Delta t_{S}$, and that $\Delta t_{S}$ can be as long as 5.1 yr. 

Our model predicts $N_{\rm TDE}/f_{\rm occ}=(10,4)$ events for $f_{\rm out}=(0.1,0.9)$, respectively. Monte Carlo simulations of TDEs occurring randomly at the rate 
$\dot N_{\rm TDE,i}$ and observed for a timescale  $\Delta t_S$ at the cadence of the NGVS confirm these basic estimates. Figure \ref{fig:fout_dist} shows the distribution of $\Delta t_{\rm TDE}$ for our GC sample at varying $f_{\rm out}$, with the weaker winds$+$disk systems outshining the steady stellar component for longer timescales because of late time accretion. 
Figure \ref{fig:tde0109} shows three examples of TDE mock detections in the Monte Carlo simulations, with the TDE light curve (the sum of the disk and wind components) compared to the steady stellar luminosity in the $u$-band. In the figure, the total time span between the first and last epoch of cluster observation is shown as the gray region.

%
\begin{figure}[!t]
    \centering 
    \includegraphics[width=0.45\textwidth]{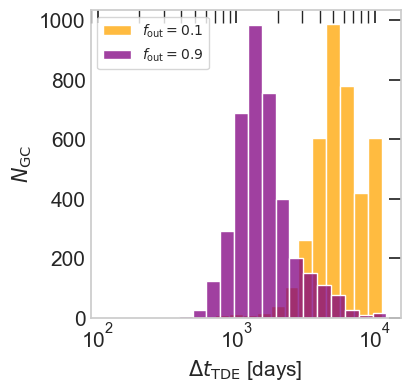}
    \caption{Distribution of $\Delta t_{\rm TDE}$ timescales (see text for details) for our sample of GCs at varying $f_{\rm out}$, with the weaker winds$+$disk systems outshinining the steady stellar component for longer timescales because of late time accretion. 
    }
    \label{fig:fout_dist}
\end{figure}
\begin{figure*}[htbp]
    \centering 
    \includegraphics[width=0.99\textwidth]{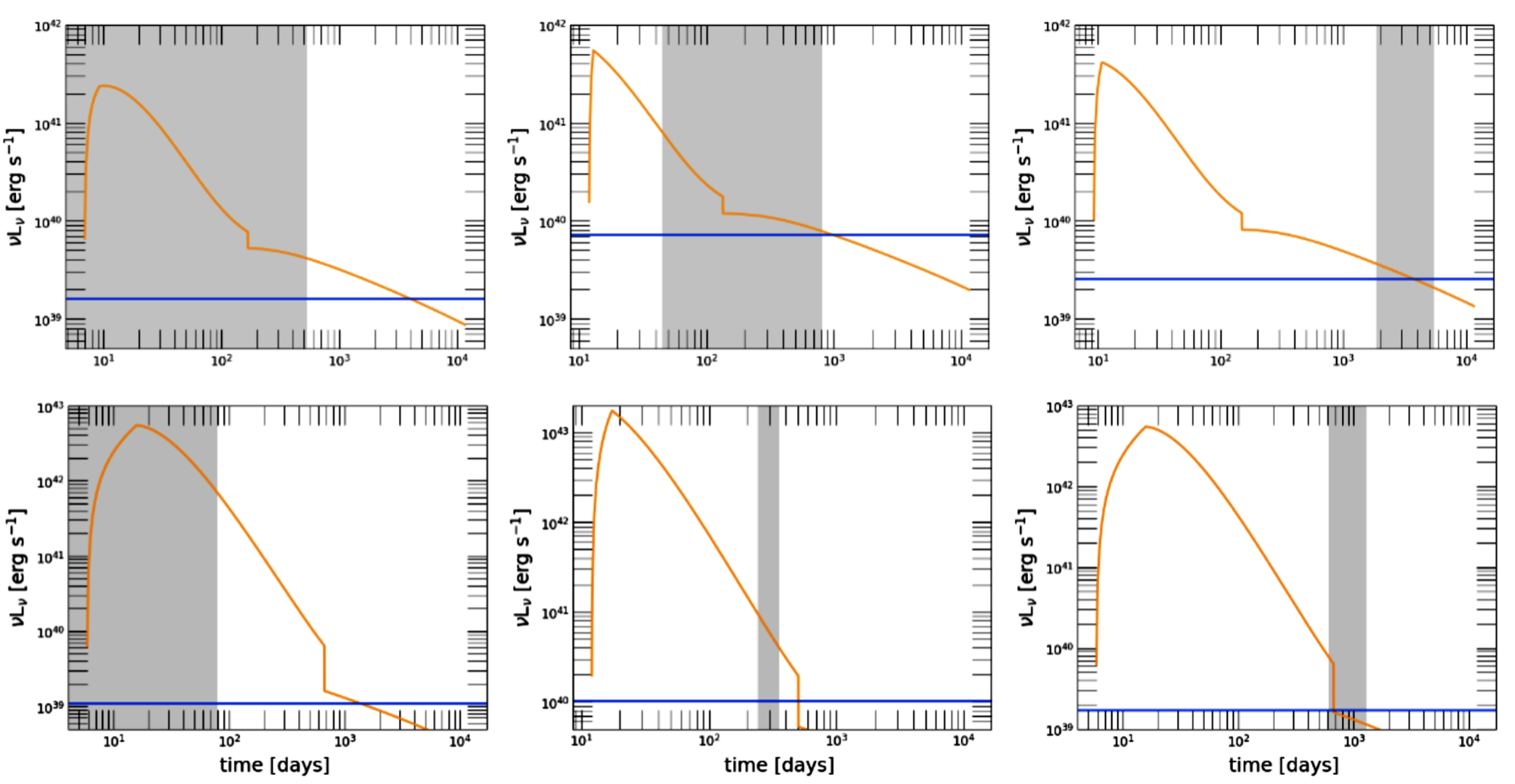}
    \caption{Three examples of TDE detections from our Monte Carlo simulations. Each light curve is compute in the $u$-band (solid orange line) for $\fout = 0.1$ (top panel) and $\fout = 0.9$ (bottom panel)
    with IMBH mass $\mbh = 0.01\,M_{\rm GC}$. The solid blue line shows the GC mean monochromatic luminosity while the gray shaded region marks the total time span between the first and last epoch of cluster observation, $\Delta t_S$ (see text for details).  From top 
    left to bottom right, the masses of the IMBHs responsible for the TDE are $1.39\times 10^4, 1.26\times 10^5,9.25\times 10^4, 3.02\times 10^4, 1.26\times 10^5,$ and $3.03\times 10^4\,\msun$.
    }
    \label{fig:tde0109}
\end{figure*}

\section{Summary and Discussion}\label{summary}

We have conducted a pilot study to search for IMBHs in dense stellar systems, one that uses TDEs as a probe
of the presence of $10^3\,\msun \lesssim \mbh \lesssim 10^{5}\,\msun$ black holes embedded in stellar cusps at the center of massive GCs. Following previous work, we have  modelled the long super-Eddington accretion phase in the slim advective disk regime together with the accompanying adiabatically expanding radiation-driven outflow. The ensuing optical/UV flare easily outshines the brightness of the (old) stellar population of globulars for a period of months to years, making TDEs triggered by  IMBHs in principle detectable in a large sample of GCs. The disruption rate of main-sequence stars as a function of black hole mass and GC properties was estimated with a simple model of loss-cone dynamics and the help of a numerical 1D Fokker-Planck approach.

Large surveys with precise multi-wavelength photometry are becoming increasingly available for public use. As an illustrative example, we have taken advantage of the NGVS, an optical-near IR imaging survey of the Virgo galaxy cluster, and of its robust sample of GCs observed  in the $u$- and $g$-band. We have checked for the presence of large photometric variations in the $u$-band induced by 
potential TDEs flares and found no obvious candidates.
Since our model predicts as many as $10\times f_{\rm occ}$ detectable events in the NGVS sample, the lack of recognizable candidates in the data implies that the fraction of GCs harboring IMBHs must be $f_{\rm occ}\lta 10\%$. This is not very constraining, as post-merger recoil kicks originated by anysotropic GW emission may make it hard for IMBHs to be retained in lower-dispersion parent clusters \citep[e.g.][]{Sedda2023}.

Naturally, better modeling of the properties of the cluster central stellar distribution and of TDE light curves and rates are all needed to gain deeper insights into the presence of IMBHs in GCs. Large samples of extra-galactic star clusters, like those  that will be assembled by the Vera C. Rubin Observatory \citep{Usher2023}, should enable significant progress in the search for TDEs in dense stellar systems. While there are considerable uncertainties in the modeling of TDE events in GCs, we hope that our pilot investigation will provide a blueprint for future searches of optical flares triggered by IMBHs.

\section*{Acknowledgements}

Support for this work was provided by NASA through grants 80NSSC21K027 and 80NSSC22K0814 (P.M.) and by a Eugene-Cota Robles Fellowship to V.T. E.B.   acknowledges support from the European Research Council (ERC) under the European Union's Horizon 2020 research and innovation program ERC-2018-CoG
under grant agreement 818691 (B~Massive), from the European Consortium for Astroparticle Theory in the form of an Exchange Travel Grant, and the European Union’s Horizon 2020 Programme under the AHEAD2020 project (grant agreement 871158). We acknowledge useful discussions on this project with Y. Feng, R. Guhathakurta, G. Lodato, and E. Rossi.

\bibliographystyle{apj}
\bibliography{references}

\begin{thebibliography}{}
\expandafter\ifx\csname natexlab\endcsname\relax\def\natexlab#1{#1}\fi

\bibitem[{{Alexander} \& {Hopman}(2009)}]{alexander2009}
{Alexander}, T., \& {Hopman}, C. 2009, \apj, 697, 1861

\bibitem[{{Arca Sedda} {et~al.}(2023){Arca Sedda}, {Kamlah}, {Spurzem},
  {Rizzuto}, {Naab}, {Giersz}, \& {Berczik}}]{Sedda2023}
{Arca Sedda}, M., {Kamlah}, A. W.~H., {Spurzem}, R., {et~al.} 2023, \mnras,
  526, 429

\bibitem[{{Bahcall} \& {Wolf}(1977)}]{bahcall1977}
{Bahcall}, J.~N., \& {Wolf}, R.~A. 1977, \apj, 216, 883

\bibitem[{{Baumgardt} \& {Hilker}(2018)}]{Baumgardt2018}
{Baumgardt}, H., \& {Hilker}, M. 2018, \mnras, 478, 1520

\bibitem[{{Baumgardt} {et~al.}(2004){Baumgardt}, {Makino}, \&
  {Ebisuzaki}}]{baumgardt2004}
{Baumgardt}, H., {Makino}, J., \& {Ebisuzaki}, T. 2004, \apj, 613, 1133

\bibitem[{{Baumgardt} {et~al.}(2019){Baumgardt}, {He}, {Sweet}, {Drinkwater},
  {Sollima}, {Hurley}, {Usher}, {Kamann}, {Dalgleish}, {Dreizler}, \&
  {Husser}}]{baumgardt2019}
{Baumgardt}, H., {He}, C., {Sweet}, S.~M., {et~al.} 2019, \mnras, 488, 5340

\bibitem[{{Binney} \& {Tremaine}(1987)}]{binney1987}
{Binney}, J., \& {Tremaine}, S. 1987, {Galactic dynamics}

\bibitem[{{Bortolas}(2022)}]{bortolas2022}
{Bortolas}, E. 2022, \mnras, 511, 2885

\bibitem[{{Bovy} {et~al.}(2011){Bovy}, {Hogg}, \& {Roweis}}]{bovy2011}
{Bovy}, J., {Hogg}, D.~W., \& {Roweis}, S.~T. 2011, Annals of Applied
  Statistics, 5, 1657

\bibitem[{{Chen} \& {Shen}(2018)}]{chen2018}
{Chen}, J.-H., \& {Shen}, R.-F. 2018, \apj, 867, 20

\bibitem[{{Coughlin} \& {Begelman}(2014)}]{coughlin2014}
{Coughlin}, E.~R., \& {Begelman}, M.~C. 2014, \apj, 781, 82

\bibitem[{{Coughlin} \& {Nixon}(2019)}]{coughlin2019}
{Coughlin}, E.~R., \& {Nixon}, C.~J. 2019, \apjl, 883, L17

\bibitem[{{Dotan} \& {Shaviv}(2011)}]{dotan2011}
{Dotan}, C., \& {Shaviv}, N.~J. 2011, \mnras, 413, 1623

\bibitem[{{Evans} \& {Kochanek}(1989)}]{evans1989}
{Evans}, C.~R., \& {Kochanek}, C.~S. 1989, \apjl, 346, L13

\bibitem[{{Feldmeier} {et~al.}(2013){Feldmeier}, {L{\"u}tzgendorf}, {Neumayer},
  {Kissler-Patig}, {Gebhardt}, {Baumgardt}, {Noyola}, {de Zeeuw}, \&
  {Jalali}}]{feldmeier2013}
{Feldmeier}, A., {L{\"u}tzgendorf}, N., {Neumayer}, N., {et~al.} 2013, \aap,
  554, A63

\bibitem[{{Ferrarese} {et~al.}(2012){Ferrarese}, {C{\^o}t{\'e}}, {Cuillandre},
  {Gwyn}, {Peng}, {MacArthur}, {Duc}, {Boselli}, {Mei}, {Erben}, {McConnachie},
  {Durrell}, {Mihos}, {Jord{\'a}n}, {Lan{\c{c}}on}, {Puzia}, {Emsellem},
  {Balogh}, {Blakeslee}, {van Waerbeke}, {Gavazzi}, {Vollmer}, {Kavelaars},
  {Woods}, {Ball}, {Boissier}, {Courteau}, {Ferriere}, {Gavazzi},
  {Hildebrandt}, {Hudelot}, {Huertas-Company}, {Liu}, {McLaughlin}, {Mellier},
  {Milkeraitis}, {Schade}, {Balkowski}, {Bournaud}, {Carlberg}, {Chapman},
  {Hoekstra}, {Peng}, {Sawicki}, {Simard}, {Taylor}, {Tully}, {van Driel},
  {Wilson}, {Burdullis}, {Mahoney}, \& {Manset}}]{ferrarese2012}
{Ferrarese}, L., {C{\^o}t{\'e}}, P., {Cuillandre}, J.-C., {et~al.} 2012, \apjs,
  200, 4

\bibitem[{{Fragione} {et~al.}(2018{\natexlab{a}}){Fragione}, {Ginsburg}, \&
  {Kocsis}}]{fragione_kicks2018}
{Fragione}, G., {Ginsburg}, I., \& {Kocsis}, B. 2018{\natexlab{a}}, \apj, 856,
  92

\bibitem[{{Fragione} {et~al.}(2018{\natexlab{b}}){Fragione}, {Leigh},
  {Ginsburg}, \& {Kocsis}}]{fragione_TDE2018}
{Fragione}, G., {Leigh}, N. W.~C., {Ginsburg}, I., \& {Kocsis}, B.
  2018{\natexlab{b}}, \apj, 867, 119

\bibitem[{{Frank} \& {Rees}(1976)}]{frank1976}
{Frank}, J., \& {Rees}, M.~J. 1976, \mnras, 176, 633

\bibitem[{{Giersz} {et~al.}(2015){Giersz}, {Leigh}, {Hypki}, {L{\"u}tzgendorf},
  \& {Askar}}]{giersz2015}
{Giersz}, M., {Leigh}, N., {Hypki}, A., {L{\"u}tzgendorf}, N., \& {Askar}, A.
  2015, \mnras, 454, 3150

\bibitem[{{Greene} {et~al.}(2020){Greene}, {Strader}, \& {Ho}}]{greene2020}
{Greene}, J.~E., {Strader}, J., \& {Ho}, L.~C. 2020, \araa, 58, 257

\bibitem[{{Guillochon} \& {Ramirez-Ruiz}(2013)}]{guillochon2013}
{Guillochon}, J., \& {Ramirez-Ruiz}, E. 2013, \apj, 767, 25

\bibitem[{{Holley-Bockelmann} {et~al.}(2008){Holley-Bockelmann},
  {G{\"u}ltekin}, {Shoemaker}, \& {Yunes}}]{holley2008}
{Holley-Bockelmann}, K., {G{\"u}ltekin}, K., {Shoemaker}, D., \& {Yunes}, N.
  2008, \apj, 686, 829

\bibitem[{{Ibata} {et~al.}(2009){Ibata}, {Bellazzini}, {Chapman},
  {Dalessandro}, {Ferraro}, {Irwin}, {Lanzoni}, {Lewis}, {Mackey}, {Miocchi},
  \& {Varghese}}]{ibata2009}
{Ibata}, R., {Bellazzini}, M., {Chapman}, S.~C., {et~al.} 2009, \apjl, 699,
  L169

\bibitem[{{Jord{\'a}n} {et~al.}(2007){Jord{\'a}n}, {McLaughlin},
  {C{\^o}t{\'e}}, {Ferrarese}, {Peng}, {Mei}, {Villegas}, {Merritt}, {Tonry},
  \& {West}}]{jordan2007}
{Jord{\'a}n}, A., {McLaughlin}, D.~E., {C{\^o}t{\'e}}, P., {et~al.} 2007,
  \apjs, 171, 101

\bibitem[{{Kamann} {et~al.}(2016){Kamann}, {Husser}, {Brinchmann}, {Emsellem},
  {Weilbacher}, {Wisotzki}, {Wendt}, {Krajnovi{\'c}}, {Roth}, {Bacon}, \&
  {Dreizler}}]{kamann2016}
{Kamann}, S., {Husser}, T.~O., {Brinchmann}, J., {et~al.} 2016, \aap, 588, A149

\bibitem[{{Kippenhahn} \& {Weigert}(1990)}]{kippen1990}
{Kippenhahn}, R., \& {Weigert}, A. 1990, {Stellar Structure and Evolution}

\bibitem[{{Kitaki} {et~al.}(2021){Kitaki}, {Mineshige}, {Ohsuga}, \&
  {Kawashima}}]{kitaki2021}
{Kitaki}, T., {Mineshige}, S., {Ohsuga}, K., \& {Kawashima}, T. 2021, \pasj,
  73, 450

\bibitem[{{Kochanek}(2016)}]{kochanek2016}
{Kochanek}, C.~S. 2016, \mnras, 461, 371

\bibitem[{{Law-Smith} {et~al.}(2020){Law-Smith}, {Coulter}, {Guillochon},
  {Mockler}, \& {Ramirez-Ruiz}}]{law-smith2020}
{Law-Smith}, J. A.~P., {Coulter}, D.~A., {Guillochon}, J., {Mockler}, B., \&
  {Ramirez-Ruiz}, E. 2020, \apj, 905, 141

\bibitem[{{Lightman} \& {Shapiro}(1977)}]{lightman1977}
{Lightman}, A.~P., \& {Shapiro}, S.~L. 1977, \apj, 211, 244

\bibitem[{{Lin} {et~al.}(2018){Lin}, {Strader}, {Carrasco}, {Page},
  {Romanowsky}, {Homan}, {Irwin}, {Remillard}, {Godet}, {Webb}, {Baumgardt},
  {Wijnands}, {Barret}, {Duc}, {Brodie}, \& {Gwyn}}]{lin2018}
{Lin}, D., {Strader}, J., {Carrasco}, E.~R., {et~al.} 2018, Nature Astronomy,
  2, 656

\bibitem[{{Lodato} {et~al.}(2009){Lodato}, {King}, \& {Pringle}}]{lodato2009}
{Lodato}, G., {King}, A.~R., \& {Pringle}, J.~E. 2009, \mnras, 392, 332

\bibitem[{{Lodato} \& {Rossi}(2011)}]{lodato2011}
{Lodato}, G., \& {Rossi}, E.~M. 2011, \mnras, 410, 359

\bibitem[{{Lu} \& {Bonnerot}(2020)}]{lu2020}
{Lu}, W., \& {Bonnerot}, C. 2020, \mnras, 492, 686

\bibitem[{{L{\"u}tzgendorf} {et~al.}(2013){L{\"u}tzgendorf}, {Kissler-Patig},
  {Neumayer}, {Baumgardt}, {Noyola}, {de Zeeuw}, {Gebhardt}, {Jalali}, \&
  {Feldmeier}}]{lutz2013}
{L{\"u}tzgendorf}, N., {Kissler-Patig}, M., {Neumayer}, N., {et~al.} 2013,
  \aap, 555, A26

\bibitem[{{Magorrian} \& {Tremaine}(1999)}]{magorrian1999}
{Magorrian}, J., \& {Tremaine}, S. 1999, \mnras, 309, 447

\bibitem[{Mei {et~al.}(2007)Mei, Blakeslee, Cote, Tonry, West, Ferrarese,
  Jordan, Peng, Anthony, \& Merritt}]{Mei_2007}
Mei, S., Blakeslee, J.~P., Cote, P., {et~al.} 2007, The Astrophysical Journal,
  655, 144

\bibitem[{{Merritt}(2004)}]{merritt2004}
{Merritt}, D. 2004, in Coevolution of Black Holes and Galaxies, ed. L.~C. {Ho},
  263

\bibitem[{{Metzger} \& {Stone}(2016)}]{metzger2016}
{Metzger}, B.~D., \& {Stone}, N.~C. 2016, \mnras, 461, 948

\bibitem[{{Miller} \& {Davies}(2012)}]{miller2012}
{Miller}, M.~C., \& {Davies}, M.~B. 2012, \apj, 755, 81

\bibitem[{{Miller} \& {Hamilton}(2002)}]{miller2002}
{Miller}, M.~C., \& {Hamilton}, D.~P. 2002, \mnras, 330, 232

\bibitem[{{Mockler} {et~al.}(2019){Mockler}, {Guillochon}, \&
  {Ramirez-Ruiz}}]{mockler2019}
{Mockler}, B., {Guillochon}, J., \& {Ramirez-Ruiz}, E. 2019, \apj, 872, 151

\bibitem[{{Mu{\~n}oz} {et~al.}(2014){Mu{\~n}oz}, {Puzia}, {Lan{\c{c}}on},
  {Peng}, {C{\^o}t{\'e}}, {Ferrarese}, {Blakeslee}, {Mei}, {Cuillandre},
  {Hudelot}, {Courteau}, {Duc}, {Balogh}, {Boselli}, {Bournaud}, {Carlberg},
  {Chapman}, {Durrell}, {Eigenthaler}, {Emsellem}, {Gavazzi}, {Gwyn},
  {Huertas-Company}, {Ilbert}, {Jord{\'a}n}, {L{\"a}sker}, {Licitra}, {Liu},
  {MacArthur}, {McConnachie}, {McCracken}, {Mellier}, {Peng}, {Raichoor},
  {Taylor}, {Tonry}, {Tully}, \& {Zhang}}]{munoz2014}
{Mu{\~n}oz}, R.~P., {Puzia}, T.~H., {Lan{\c{c}}on}, A., {et~al.} 2014, \apjs,
  210, 4

\bibitem[{{Nguyen} {et~al.}(2019){Nguyen}, {Seth}, {Neumayer}, {Iguchi},
  {Cappellari}, {Strader}, {Chomiuk}, {Tremou}, {Pacucci}, {Nakanishi},
  {Bahramian}, {Nguyen}, {den Brok}, {Ahn}, {Voggel}, {Kacharov}, {Tsukui},
  {Ly}, {Dumont}, \& {Pechetti}}]{nguyen2019}
{Nguyen}, D.~D., {Seth}, A.~C., {Neumayer}, N., {et~al.} 2019, \apj, 872, 104

\bibitem[{{Ohsuga} {et~al.}(2005){Ohsuga}, {Mori}, {Nakamoto}, \&
  {Mineshige}}]{ohsuga2005}
{Ohsuga}, K., {Mori}, M., {Nakamoto}, T., \& {Mineshige}, S. 2005, \apj, 628,
  368

\bibitem[{{Portegies Zwart} \& {McMillan}(2002)}]{zwart2002}
{Portegies Zwart}, S.~F., \& {McMillan}, S. L.~W. 2002, \apj, 576, 899

\bibitem[{{Ramirez-Ruiz} \& {Rosswog}(2009)}]{ramirez2009}
{Ramirez-Ruiz}, E., \& {Rosswog}, S. 2009, \apjl, 697, L77

\bibitem[{{Rees}(1988)}]{rees1988}
{Rees}, M.~J. 1988, \nat, 333, 523

\bibitem[{{Rossi} {et~al.}(2021){Rossi}, {Stone}, {Law-Smith}, {Macleod},
  {Lodato}, {Dai}, \& {Mandel}}]{rossi2021}
{Rossi}, E.~M., {Stone}, N.~C., {Law-Smith}, J.~A.~P., {et~al.} 2021, \ssr,
  217, 40

\bibitem[{{Roth} {et~al.}(2020){Roth}, {Rossi}, {Krolik}, {Piran}, {Mockler},
  \& {Kasen}}]{roth2020}
{Roth}, N., {Rossi}, E.~M., {Krolik}, J., {et~al.} 2020, \ssr, 216, 114

\bibitem[{{Ryu} {et~al.}(2020){Ryu}, {Krolik}, \& {Piran}}]{ryu2020}
{Ryu}, T., {Krolik}, J., \& {Piran}, T. 2020, \apj, 904, 73

\bibitem[{{Spitzer}(1987)}]{spitzer1987}
{Spitzer}, L. 1987, {Dynamical evolution of globular clusters}

\bibitem[{{Stone} \& {Metzger}(2016)}]{stone2016}
{Stone}, N.~C., \& {Metzger}, B.~D. 2016, \mnras, 455, 859

\bibitem[{{Strubbe} \& {Quataert}(2009)}]{strubbe2009}
{Strubbe}, L.~E., \& {Quataert}, E. 2009, \mnras, 400, 2070

\bibitem[{{Strubbe} \& {Quataert}(2011)}]{strubbe2011}
---. 2011, \mnras, 415, 168

\bibitem[{{Syer} \& {Ulmer}(1999)}]{syer1999}
{Syer}, D., \& {Ulmer}, A. 1999, \mnras, 306, 35

\bibitem[{{Tremou} {et~al.}(2018){Tremou}, {Strader}, {Chomiuk}, {Shishkovsky},
  {Maccarone}, {Miller-Jones}, {Tudor}, {Heinke}, {Sivakoff}, {Seth}, \&
  {Noyola}}]{tremou2018}
{Tremou}, E., {Strader}, J., {Chomiuk}, L., {et~al.} 2018, \apj, 862, 16

\bibitem[{{Usher} {et~al.}(2023){Usher}, {Dage}, {Girardi}, {Barmby},
  {Bonatto}, {Chies-Santos}, {Clarkson}, {G{\'o}mez Camus}, {Hartmann},
  {Ferguson}, {Pieres}, {Prisinzano}, {Rhode}, {Rich}, {Ripepi}, {Santiago},
  {Stassun}, {Street}, {Szab{\'o}}, {Venuti}, {Zaggia}, {Canossa}, {Floriano},
  {Lopes}, {Miranda}, {Oliveira}, {Reina-Campos}, {Roman-Lopes}, \&
  {Sobeck}}]{Usher2023}
{Usher}, C., {Dage}, K.~C., {Girardi}, L., {et~al.} 2023, \pasp, 135, 074201

\bibitem[{{Vasiliev}(2017)}]{vasiliev2017}
{Vasiliev}, E. 2017, \apj, 848, 10

\bibitem[{{Vitral} \& {Mamon}(2021)}]{vitral2021}
{Vitral}, E., \& {Mamon}, G.~A. 2021, \aap, 646, A63

\bibitem[{{Wang} \& {Merritt}(2004)}]{wang2004}
{Wang}, J., \& {Merritt}, D. 2004, \apj, 600, 149

\bibitem[{{Webb} {et~al.}(2012){Webb}, {Cseh}, {Lenc}, {Godet}, {Barret},
  {Corbel}, {Farrell}, {Fender}, {Gehrels}, \& {Heywood}}]{webb2012}
{Webb}, N., {Cseh}, D., {Lenc}, E., {et~al.} 2012, Science, 337, 554

\bibitem[{{Wen} {et~al.}(2021){Wen}, {Jonker}, {Stone}, \&
  {Zabludoff}}]{wen2021}
{Wen}, S., {Jonker}, P.~G., {Stone}, N.~C., \& {Zabludoff}, A.~I. 2021, \apj,
  918, 46

\bibitem[{{Wu} {et~al.}(2018){Wu}, {Coughlin}, \& {Nixon}}]{wu2018}
{Wu}, S., {Coughlin}, E.~R., \& {Nixon}, C. 2018, \mnras, 478, 3016

\bibitem[{{Zhao}(1996)}]{zhao1996}
{Zhao}, H. 1996, \mnras, 278, 488

\bibitem[{{Zocchi} {et~al.}(2019){Zocchi}, {Gieles}, \&
  {H{\'e}nault-Brunet}}]{zocchi2019}
{Zocchi}, A., {Gieles}, M., \& {H{\'e}nault-Brunet}, V. 2019, \mnras, 482, 4713

\end{thebibliography}
\end{document}